\documentclass[conference]{IEEEtran}

\usepackage{balance}
\usepackage{cite}

\ifCLASSINFOpdf
  \usepackage[pdftex]{graphicx}
  \usepackage{caption2}
\else
\fi

\usepackage[cmex10]{amsmath}

\usepackage{indentfirst}

\usepackage{array}

\usepackage{tikz}
\usepackage{textcomp}
\usepackage{hyperref}

\newcommand\copyrighttext{%
	\footnotesize \textcopyright 2024 IEEE. Personal use of this material is permitted.
	Permission from IEEE must be obtained for all other uses, in any current or future
	media, including reprinting/republishing this material for advertising or promotional
	purposes, creating new collective works, for resale or redistribution to servers or
	lists, or reuse of any copyrighted component of this work in other works.}
\newcommand\copyrightnotice{%
	\begin{tikzpicture}[remember picture,overlay]
	\node[anchor=south,yshift=10pt] at (current page.south) {\fbox{\parbox{\dimexpr\textwidth-\fboxsep-\fboxrule\relax}{\copyrighttext}}};
	\end{tikzpicture}%
}

\begin{document}
\title{Low-latency Symbol-Synchronous Communication for Multi-hop Sensor Networks}

\author{\IEEEauthorblockN{
                Xinlei Liu\IEEEauthorrefmark{1},
                Andrey Belogaev\IEEEauthorrefmark{1},
                Jonathan Oostvogels\IEEEauthorrefmark{2},
                Bingwu Fang\IEEEauthorrefmark{2},\\
                Danny Hughes\IEEEauthorrefmark{2}, 
                Maarten Weyn\IEEEauthorrefmark{1},
                and Jeroen Famaey\IEEEauthorrefmark{1}}
\IEEEauthorblockA{\IEEEauthorrefmark{1}IDLab, University of Antwerp -- imec, Belgium\\
        Email: \{Xinlei.Liu, Andrey.Belogaev, Jeroen.Famaey\}@uantwerpen.be}
\IEEEauthorblockA{\IEEEauthorrefmark{2}DistriNet, KU Leuven
\\
        Email: \{Jonathan.Oostvogels, Bingwu.Fang,  Danny.Hughes\}@kuleuven.be}
}

\maketitle
\copyrightnotice

\begin{abstract}
Wireless sensor networks~(WSNs) have received great interest due to their scalability, energy efficiency, and low-cost deployment. By utilizing multi-hop communication, WSNs can cover a wide area using low transmission power without the need for any communication infrastructure. Traditionally, WSNs rely on store-and-forward routing protocols and Time Division Multiple Access~(TDMA)-based schedules that avoid interference between different wireless nodes. However, emerging challenging scenarios, such as the industrial Internet of Things~(IoT) and robotic swarms, impose strict latency and reliability requirements, which traditional approaches cannot fulfill. In this paper, we propose a novel symbol-synchronous transmission design that provides reliable low-latency communication with a reasonable data rate on classical sub-6GHz RF frequency bands~(e.g., the $2.4$~GHz ISM band). Instead of avoiding overlapping transmissions, the proposed scheme benefits from concurrent transmissions. Using simulation in MATLAB, we prove that the proposed design allows achieving a wire-like delay of $5$~ms for a 512-bit packet over multiple hops with only a $0.3$\% latency increase per extra hop and a low bit error rate~(BER) of $0.04$\%. Compared to similar state-of-the-art approaches it can achieve a significantly higher data rate of $100$~kbps, which is expected to increase further with future improvements of the system.

\end{abstract}

\IEEEpeerreviewmaketitle

\section{Introduction}
The emergence of Internet of Things (IoT) technologies, which allow low-power embedded devices to communicate over the Internet, brought innovation in a wide range of fields, such as smart home, industry automation, and healthcare~\cite{10.1007/978-3-030-29407-6_20}. In addition, in the face of increasing demands from industrial applications, emerging IoT networks are demanded to achieve ultra-high reliability and low latency communication to meet diverse requirements from industry and market~\cite{electronics8090981}. With the revolution of IoT, WSNs have become an integral part of IoT networks to transfer sensed data~\cite{gupta2023anchor}. In principle, multi-hop networks rely on relaying technologies, i.e., nodes relay information to their neighboring nodes within a short range with small transmission power~\cite{badis2015modeling}. Benefited by multi-hop architecture, the IoT networks can be deployed in larger areas and cover a wider propagation range with low expenditure on network construction. Furthermore, compared to centralized networks, they are easier and more flexible to scale and deploy in areas that are deficient in infrastructure.

However, the problem of end-to-end latency reduction in multi-hop networks is attracting much interest from researchers from academia and industry. As for various latency-sensitive networks such as real-time control systems and robotic swarms, it is expected that the wireless network can attain wire-like performance in terms of latency and reliability~\cite{coletti2024real}. Multi-hop routing protocols for WSNs, such as Routing Protocol for Low-Power and Lossy Networks~(RPL) and its variants, have been designed to support various resource-constrained applications in IoT~\cite{7347997}. One of the main challenges of decreasing WSN latency is the use of the store-and-forward routing and/or TDMA, which cause significantly larger latency compared to wired alternatives, such as Controller Area Network~(CAN)~\cite{8368987,ben2020controller}. The problem becomes even more challenging in a broadcast scenario when the packet should be delivered to every node in the network. In this case, a completely interference-free broadcast schedule design is an NP-complete problem~\cite{ferrari2011efficient}. As a promising solution for this problem, a transmission paradigm called \emph{synchronous transmission} is being researched~\cite{oostvogels2020zero}. It challenges the view that packet overlaps must be avoided as in traditional transmission. Instead, it argues that the received signal redundancy can also provide useful information in certain circumstances~\cite{zimmerling2020synchronous}.

In this paper, we propose a solution for packet flooding in WSNs based on symbol-synchronous transmission in which nodes relay a message symbol by symbol. Here with flood, we mean that a message is disseminated from an initiator node to all other nodes in the network. We evaluate its performance by means of simulation using MATLAB. The rest of the paper is organized as follows. 
In Section~\ref{sec:background}, we consider state-of-the-art solutions for WSNs.
In Section~\ref{sec:design}, we describe the design of the proposed packet flooding approach based on symbol-synchronous transmissions.
In Section~\ref{sec:evaluation}, we evaluate the network performance using simulation in MATLAB.
Finally, in Section~\ref{sec:conclusion}, we conclude the paper.

\section{State of the art}
\label{sec:background}

There are multiple solutions proposed in the literature for packet flooding in wireless multi-hop networks. Glossy is a typical network flooding architecture targeting one-to-all communication across a WSN~\cite{ferrari2011efficient}. Glossy proposes to design a time-synchronization network to exploit the constructive interference for decoding the packet successfully. Blueflood is another packet synchronous dissemination protocol performing similarly to Glossy, but it is based on a multi-hop Bluetooth mesh network~\cite{nahas2021blueflood}. In addition, a routing scheme called interference coordinated routing~(ICR) is designed to make use of constructive interference and concurrent transmission to decrease the latency~\cite{9312182}. Although all the above protocols benefit from synchronous transmissions, they are still based on store-and-forward, which forces every relay node to wait before it receives and decodes all symbols included in a packet before it can relay this packet. This waiting time imposes a negative impact on the end-to-end latency of a multi-hop network since the latency linearly increases by the whole packet transmission time with each hop. For instance, in Glossy, for a 64-bit packet, the end-to-end latency increases more than twice from $0.81$ms to $1.77$ms as the number of hops increases from 3 to 5~\cite{ferrari2011efficient}.

To overcome such poor scaling behavior, a technology supporting to forward data symbol-by-symbol instead of packet-by-packet is proposed, to avoid a linear increase of the latency as a function of a number of hops. When the node receives and decodes the symbol, it instantly relays this symbol without waiting for the reception of the full packet.
Zippy is a transmission protocol for one-to-all communication that exploits symbol-by-symbol transmissions. Its advance is to design an asynchronous wake-up strategy for every node and exploit a simple On-off keying~(OOK) transmitter to achieve per-hop synchronous transmission~\cite{sutton2015zippy}. The Zippy network encodes every symbol using repetition codes and uses carrier frequency randomization to avoid destructive interference. It achieves deterministic performance with an end-to-end latency of tens of microseconds and low power consumption. However, Zippy only performs well in a network with a maximum of two hops, while its performance crucially degrades for the many-hop network and long-distance communication. Its scalability is also limited due to the usage of different carrier frequencies. Zero-Wire~\cite{oostvogels2020zero} proposes a novel concept of symbol-synchronous transmission bus. However, it relies on optical frequencies, which provide very limited communication distance. Besides, the measured data rate is lower than $20$~kbps.

Inspired by the state-of-art performance of end-to-end latency of synchronous transmission, in this paper, we propose a symbol-synchronous transmission design for wireless multi-hop networks. Different from the state-of-the-art solutions, we aim to develop a multi-hop protocol that provides low end-to-end latency and is suitable for long-distance communication based on radio frequency~(RF) transmissions. Specifically, we propose to use widely available unlicensed RF bands (e.g., the $2.4$~GHz ISM band) to broadcast information over the network. Unlike the optical band used in Zero-Wire~\cite{oostvogels2020zero}, RF bands offer multiple advantages, such as a larger coverage range and better performance in non-line-of-sight conditions. Similar to Zippy and Zero-Wire~\cite{oostvogels2020zero,sutton2015zippy}, we rely on a low-complexity OOK transmitter considering the advantage that it easily handles the interference from multiple simultaneous retransmissions of the same symbol.

\section{Network Design}
\label{sec:design}

\subsection{General architecture}
In this paper, we focus on low-latency wireless communication over a multi-hop WSN. We consider the scenario where one initiator aims to transmit data of fixed size to all other nodes in the network with high reliability and low latency. All nodes cooperatively forward the data through the network in a symbol synchronous way, allowing those further away from the source to receive it over multiple hops.

\begin{figure}[!t]
\centering
\includegraphics[width=3.5in]{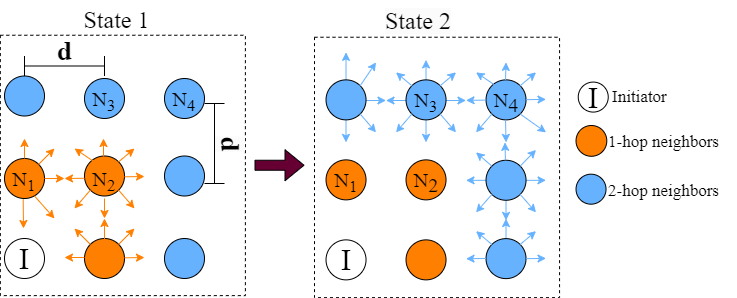}
\caption{Network topology}
\label{pic1}
\end{figure}

The proposed symbol-synchronous transmission way can be applied to various network topologies. To clearly explain the concept of the network and easily analyze experimental results, in this paper, we simplify the network topology to a grid-like topology as shown in Figure~\ref{pic1}. Figure~\ref{pic1} illustrates a static grid topology where nodes are located in a regular lattice with a fixed grid distance $d$. We assume there are $N$ nodes including one initiator. All nodes use the same OOK modulation scheme and frequency band. Considering the aim of low latency, we design a new symbol synchronous transmission strategy for the considered network. Its difference from the traditional store-and-forward transmission is that the latency is not severely and negatively affected by the number of hops. Except for the initiator, every node in the network works as a transceiver. It plays two roles, which are to receive the signal and relay it. The logic of the designed strategy is that once the node receives a symbol and decodes it, it instantly broadcasts this symbol without waiting to receive the whole packet. For simplicity, Figure~\ref{pic1} assumes that the transmission range allows only one-hop neighbors (including diagonal ones) to successfully decode symbols. As Figure~\ref{pic1} shows, once the one-hop neighbors of the initiator, such as $N_1$ and $N_2$, successfully decode the symbol, they relay it. The node $N_4$ successfully receives the symbol from the node $N_2$, while the node $N_3$ receives it from concurrent transmissions of $N_1$ and $N_2$. After that, the two-hop neighbors relay the symbol thus allowing its propagation further in the network. The performance evaluation shows that with this strategy the packets can reach all the nodes with low latency, while using low transmission power.

\subsection{Pulse-based OOK modulation scheme}
\begin{figure}[!t]
\centering
\includegraphics[width=2.5in]{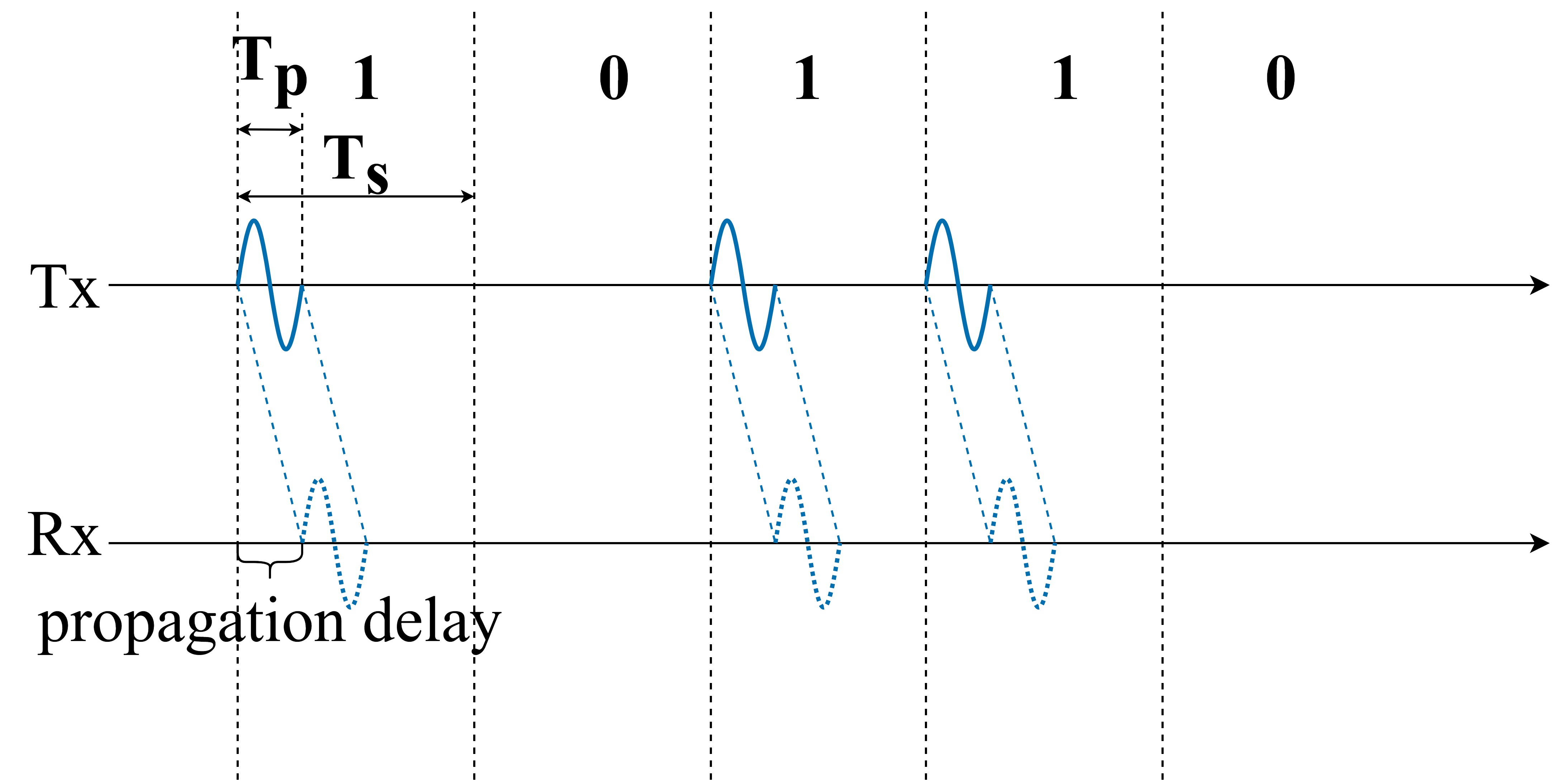}
\caption{Pulse-based OOK modulation scheme}
\label{pic2}
\end{figure}

We select pulse-based OOK as a modulation scheme. This modulation scheme is easy to implement on hardware, especially for resource-constrained devices due to its low-complexity demodulation at low power. The inter-symbol duration is $T_s$. As Figure~\ref{pic2} illustrates, when the symbol 1 is sent, the transmitter sends a short pulse with duration $T_p \ll T_s$. Otherwise, if the symbol 0 is sent, the transmitter keeps silent during $T_s$.

\subsection{Symbol-synchronous transmission strategy}
\begin{figure}[!t]
\centering
\includegraphics[width=2.5in]{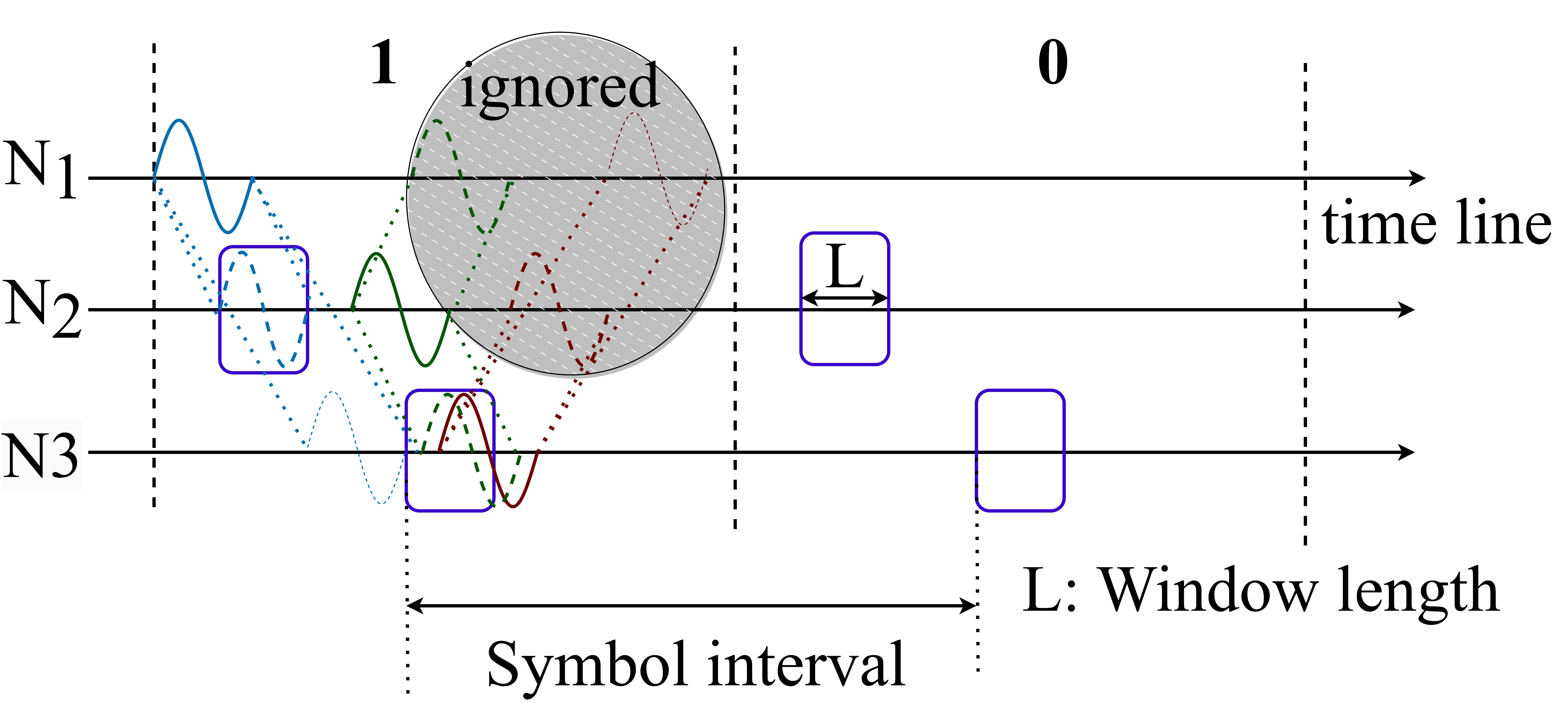}
\caption{Symbol-synchronous transmission }
\label{pic3}
\end{figure}

To achieve low-latency wireless communication, we design a novel symbol-synchronous transmission strategy, i.e., transmitting data symbol by symbol. Let us define the relay time $r$ as the sum of signal propagation delay, symbol detection delay, and processing delay. We assume that the relay time is much less than the symbol duration $T_s$ so that every node finishes retransmitting a symbol before it starts detecting the next symbol and it does not overhear retransmissions of the previous symbol while receiving the next symbol. In our analysis, we assume that the hardware processing delays are too small to impose a significant impact on the overall latency of the network, and thus are neglected. The end-to-end latency $D$ of the network is constrained by
\begin{equation}
    (n-1)T_s \leq D \leq (n-1)T_s+rh \qquad r \ll T_s,
\end{equation}
where $h$ is the number of hops from the initiator to the most distant node in the network, and $n$ is the number of transmitted bits in a packet. Using this transmission pattern, the increase of one hop for a multi-hop network only adds $r$ to latency.
In contrast, the end-to-end latency of store-and-forward transmission increases per additional hop by the full packet transmission time and additional waiting time related to processing and channel access at the MAC layer. The transmission logic is illustrated in Figure~\ref{pic3}, if the node $N_1$ receives symbol 1, it relays this symbol and sends a short pulse to nodes $N_2$ and $N_3$. After a propagation delay, the nodes $N_2$ and $N_3$ respectively receive the relayed symbol at different moments caused by different distances from $N_1$. Then a detection procedure will be conducted to decode the received symbol. Once the node $N_2$ successfully decodes the signals, it will modulate the decoded symbol based on the pulse-based OOK scheme and relay the modulated signals to $N_1$ and $N_3$. The node $N_3$ goes through a similar procedure. After a decoding process, $N_3$ will relay the decoded symbol to $N_1$ and $N_2$. On the other hand, when the node $N_1$ detects the symbol 0, it keeps silent. Similarly, the nodes $N_2$ and $N_3$ also keep silent during the second $T_s$ interval, corresponding to the symbol 0.

\subsection{Detector logic}
This section provides some details about the detection procedures based on simulation. For the initial synchronization and detection of a new transmission, it is assumed that each transmission starts with a bit 1 that is used as a synchronization preamble in this paper. To improve robustness and avoid detecting fake symbols, a more complex preamble could be used as well. This is left for future work. As the inter-symbol duration $T_s$ is fixed and assumed known to all nodes, a node can use the detection time of the preamble to re-synchronize its receiver. Additionally, a wake-up receiver~\cite{sutton2015zippy} can be considered to prevent nodes from continuous listening for preamble detection. Once a receiver wakes up, it expects to receive the first symbol within a time interval of duration $T_s$ starting from the preamble detection. Then, every next symbol will be detected in the corresponding time interval of duration $T_s$, which we call the \emph{symbol interval}. 

In the proposed system, every node will likely receive multiple copies of each symbol representing a 1 bit, as shown in the first $T_s$ time interval depicted in Figure~\ref{pic3}. During a symbol duration, these relayed signals can be received at different moments leading to multiple detections of the same symbol. It is not only unnecessary but also a waste of energy, because a longer detection time results in higher power consumption. 
To address this problem, we introduce a window with length $L$ for detection. As such, a node will only go into reception mode for a time window $L$ at the start of each symbol interval~(cf., Figure~\ref{pic3}). Once the detector detects the symbol 1, it will instantly relay it and then sleep until the next symbol interval. Using this method, we can guarantee that in every time slot $T_s$, only one symbol is detected while its relayed copies are ignored.

\begin{figure}[!ht]
\centering
\includegraphics[width=2.35in]{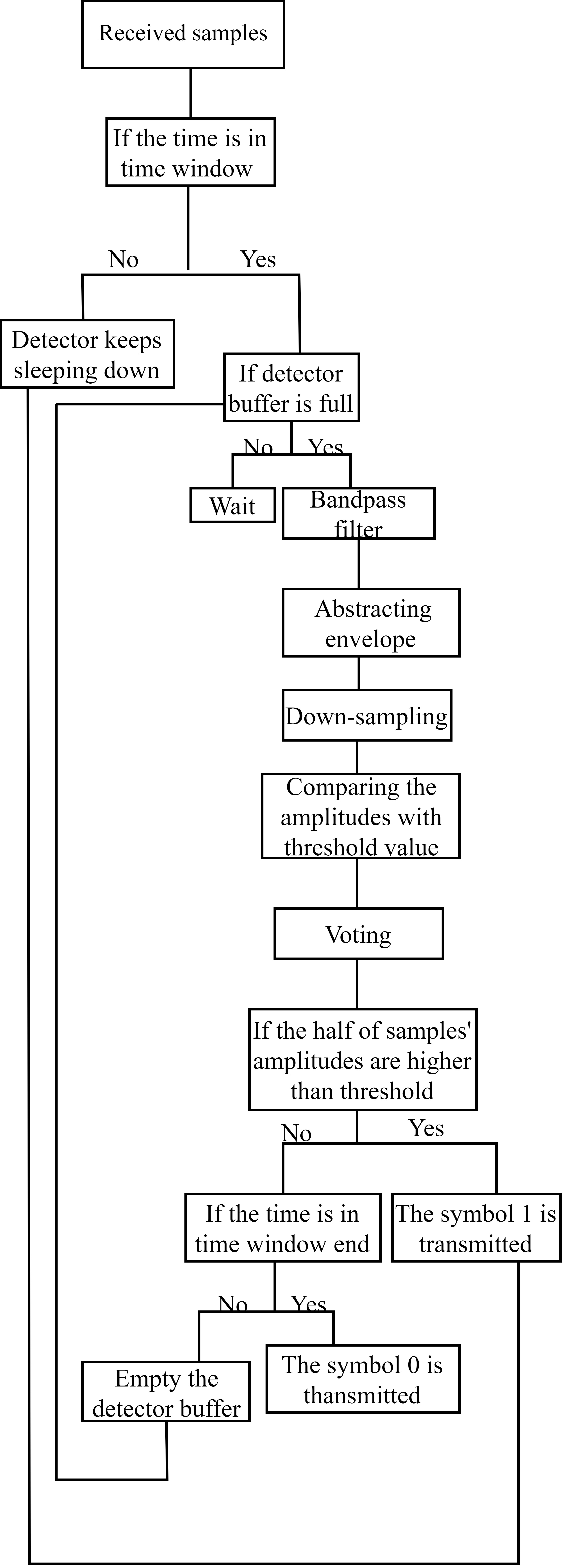}
\caption{Detection scheme }
\label{pic4}
\end{figure}    

Considering the signal overlapping due to various relay delays, the designed detection process is conducted several times during the time window $L$, improving the detection accuracy. And then we introduce a detector buffer to trigger every detection. Specifically, the received sampled signal from the transmitter is stored in a buffer until the buffer is full. After that, all stored samples in this buffer are used to make a decision by filtering the stored samples in the current buffer using a bandpass filter with a bandwidth $f_p{}_a{}_s{}_s$, which limits the noise level. After that, the envelopes of stored samples are abstracted. Because OOK modulation is essentially an amplitude modulation, an envelope detector is an efficient way to detect amplitudes disturbed by noise.
After the down-sampling of the envelope, the samples go through the amplitude comparator, and their amplitudes are compared with a fixed threshold value determined by the receiver sensitivity. Finally, we use a voting strategy to decide which symbol is decoded. Specifically, if over half of their amplitudes are higher than the threshold value, we assume the symbol 1 is detected and the detection process is terminated until the next window time starts. Otherwise, the current buffer is emptied and starts storing the new samples for the next detection. If the detector still cannot detect a pulse at the end of the window time, we decide that the symbol 0 is detected. 

\section{System evaluation}
\label{sec:evaluation}
This section describes in detail the network simulation and provides numerical results proving the network's advanced performance on end-to-end latency and reliability referring to the network BER. 
\subsection{Simulation setup}
We simulate a WSN in MATLAB. At the transmitter side, we simulate a carrier wave at $2.4$~GHz and modulate the information bits according to a pulse-based OOK scheme.  The transmit power is set as $0$~dBm, which is reasonable for resource-constrained IoT devices. The pulse duration for a symbol 1 lasts $0.2\mu\text{s}$, while the symbol interval duration $T_s$ is $10\mu\text{s}$. Hence, the data is sent with a bit rate of $100$~kbps. In case a symbol 0 is transmitted, the transmitter keeps silent during the whole slot $T_s$. 
In addition to the modulation, we also apply a pulse-shaping filter into the transmitter to further limit the bandwidth of the transmitted signal to $10$~MHz.  
Regarding the network topology simulation, the nodes are positioned in a square area of $4~\text{km}^2$ surrounded by walls where the distance $d$ between horizontal and vertical neighbors is varied across different experiments. And, the initiator is fixed in the corner. 
Between all pairs of nodes, the ray-tracing channel model~\cite{9277601} is deployed to calculate the signal attenuation and phase shift. 

To simulate the analog signal transmission in MATLAB, we use a higher frequency of $9.6$~GHz at the transmitter to sample the modulated signals. At the receiver side, we set the buffer size to $1000$~samples to trigger every detection. A non-coherent detector is simulated consisting of a bandpass filter, envelop detector, down-sampler, amplitude comparator, and voting block. Before the amplitude comparator, a down-sampler with the frequency of $96$~MHz is applied. In the simulation, the window duration time $L$ is set as $1.875\mu\text{s}$ to control the detector status so that in every window duration interval, the detection process is conducted $18$ times. 

The bandpass filter with $10$~MHz is added to limit the noise level of the receiver. We simulate the receiver noise with a band-limited Additive White Gaussian Noise~(AWGN) of the power $-103$~dBm corresponding to the room temperature $290$~K. Additionally, we set the receiver noise figure to $5$~dB. The parameters' values are listed in Table~\ref{tab:simple_table}. 
\begin{table}[!t]
  \centering
  \caption{Parameter settings in simulation}
  \label{tab:simple_table}
  \begin{tabular}{c|c}
    \hline
    \textbf{Parameters} & \textbf{Values(units)} \\
    \hline
    Data rate & $100$~kbps\\
    
    Network area & $4~\text{km}^2$\\
    Sampling frequency of transmitter & $9.6$~GHz\\
    Carrier frequency & $2.4$~GHz\\
    Pulse duration & $0.2\mu$s\\
    Transmitter power & $0$~dBm\\
    Noise power & $-103$~dBm\\
    Noise figure & $5$~dB \\
    Receiver sensitivity & $-90$~dBm\\
    Window time & $1.875\mu$s\\
    Buffer size & $1000$ samples \\
    Number of times for detection & $18$\\
    Sampling frequency of receiver & $96$~MHz\\
    Bandwidth & $10$~MHz\\
     \hline
  \end{tabular}
\end{table}
\subsection{Performance evaluation}
This section describes and discusses the experimental results. We focus on the end-to-end latency and BER indicating network reliability. The end-to-end latency means the delay between the time when the initiator sends the first symbol of a packet and the time at which all destination nodes receive and decode the packet's last symbol. The BER is represented by the average probability of incorrectly decoding a bit across all nodes in the network. The present results do not consider retransmissions or the use of error correction codes, which could improve reliability at the cost of data rate. The integration of such strategies is left for future work. Firstly, we explore the network BER as a function of different grid distances. We place $16$ nodes~($4$-by-$4$) including an initiator in a closed square area and vary the grid distance from $50$~m to $200$~m. We randomly generate and send $100$ packets consisting of $64$~bits each.
\begin{figure}[!t]
\centering
\includegraphics[width=2.5in]{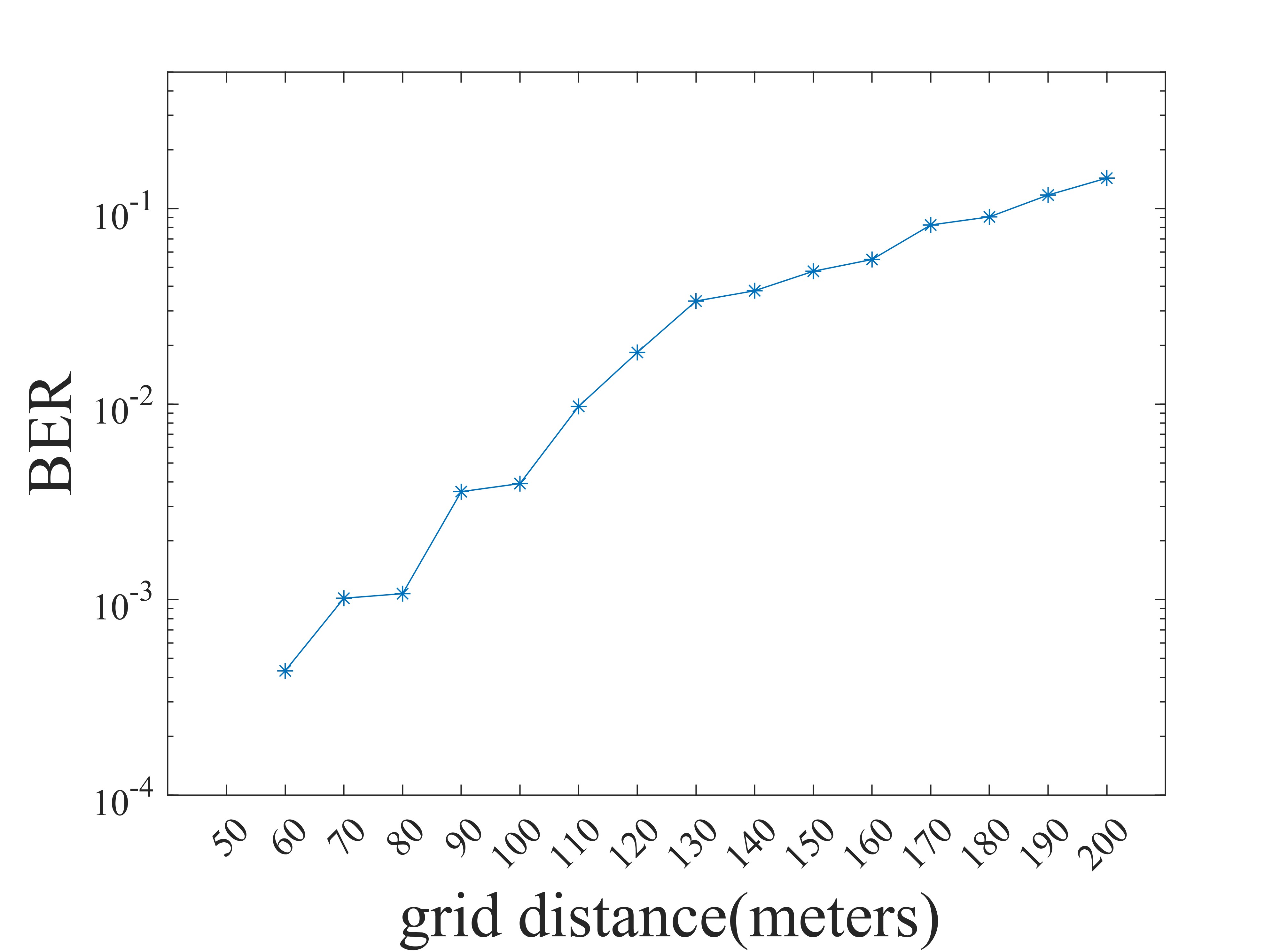}
\caption{Average BER for various grid distances.}
\label{pic5}
\end{figure}
The results are depicted in Figure~\ref{pic5}. As expected, BER increases with the grid distance. As the grid distance i.e., the distance between horizontal and vertical neighbors in the grid, increases, propagation loss will increase, due to longer links, but also the number of hops to reach further nodes will increase. We can see the network BER is lower than $1$\% when the grid distance is below $100$~m. Particularly, when the grid distance is $50$~m, we cannot see any errors for 100 packets. On the grid distance of $60$~m, the BER is $0.04$\%. When the grid distance is over $100$~m, the network BER increases sharply, because the transmission power of $0$~dBm is too low to support decoding with high accuracy even for the nearest nodes from the initiator~(i.e., one-hop nodes).

\begin{figure}[!t]
\centering
\includegraphics[width=2.5in]{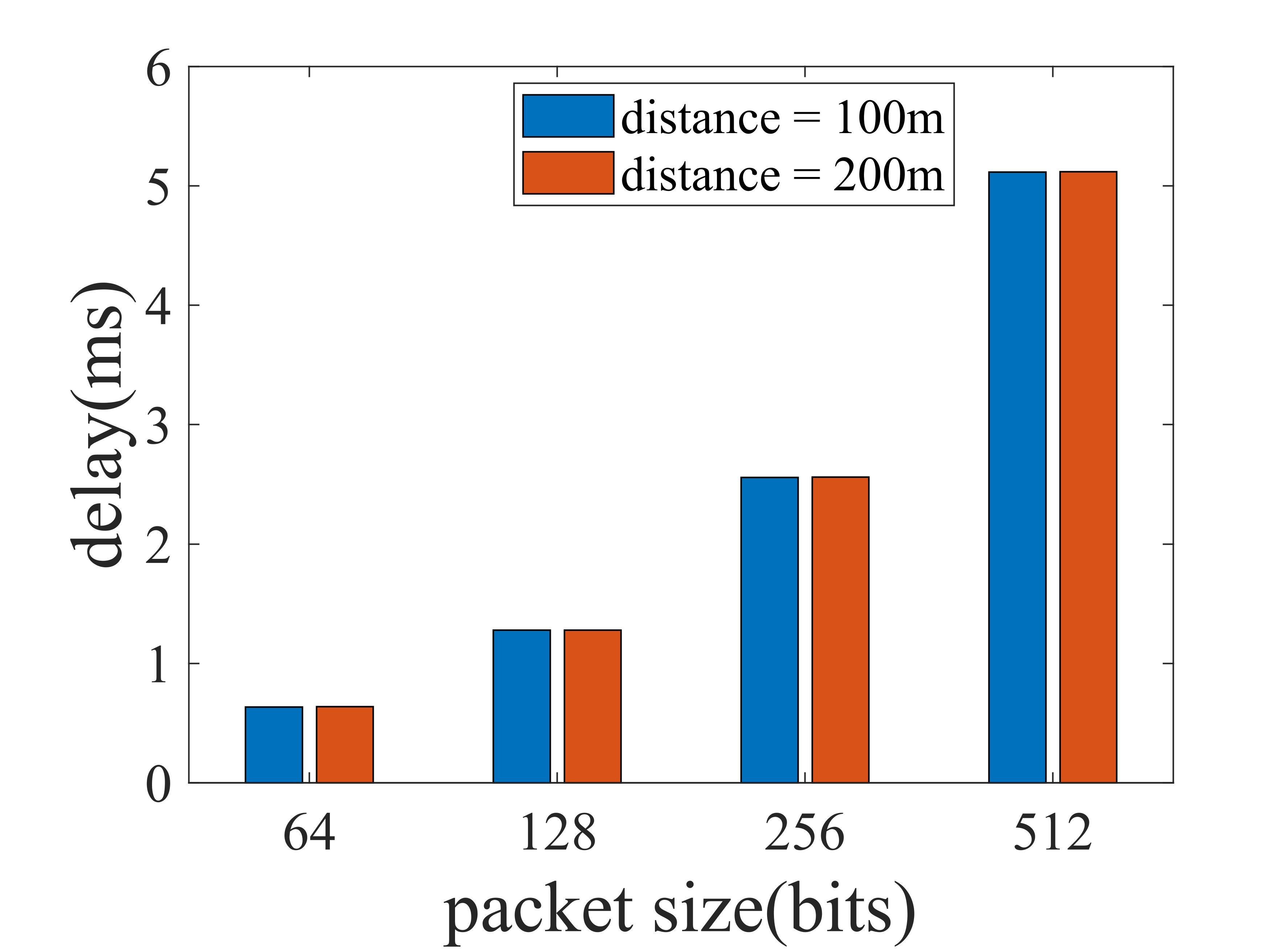}
\caption{Network end-to-end latency for grid distance $100$~m and $200$~m.}
\label{pic7}
\end{figure}

In addition, we evaluate our network for end-to-end latency, for packet sizes equal $\{64, 128, 256, 512\}$~bits across the $16$-node network with grid distance ranging from $50$~m to $200$~m. As a function of packet size, the latency almost linearly increases as shown in Figure~\ref{pic7}. According to Figure~\ref{pic7}, our network end-to-end latency is in the order of a few milliseconds. Specifically, the average latency to transmit a 512-bit packet is around 5 milliseconds. For a small packet of 64 bits, the average latency is only around 0.63 milliseconds. Moreover, the latency does not seriously increase by grid distance increase. The latency increase caused by distance increase from $100$~m to $200$~m is of orders of a few microseconds as shown in Figure~\ref{pic7}. The slight increase is due to the increased propagation delay and is negligible.

\begin{figure}[!t]
\centering
\includegraphics[width= 2.5in]{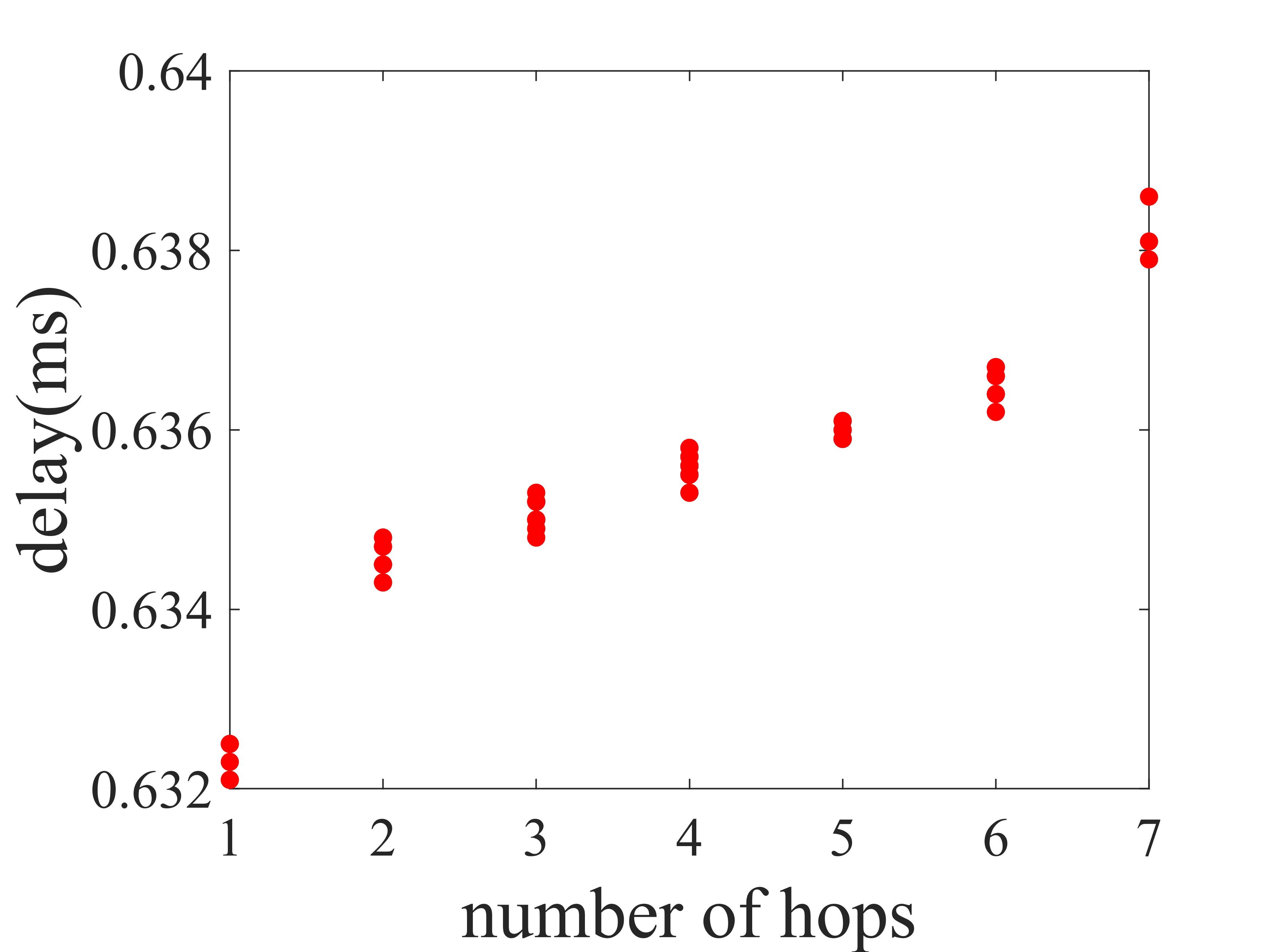}
\caption{Effect of the number of hops on latency for $64$ bits packet and grid distance $100$~m.}
\label{pic8}
\end{figure}

To explore the latency behavior as a function of the number of hops, which is the main advantage of symbol synchronous transmission compared to store-and-forward, we perform another experiment to expand our network into $64$ nodes~($8$-by-$8$), using the same network topology as shown in Figure~\ref{pic1} with a grid distance of $100$~m. Figure~\ref{pic8} depicts the latency of each node depending on their distance from the initiator in terms of the number of hops.
While the number of hops increases as a function of the distance, the figure shows that the increase is limited to about $2\mu\text{s}$ per hop~(i.e., the pulse duration of $0.2\mu\text{s}$). This results in a latency increase of about $\sim 0.3$\% per hop. This is an immense improvement compared to store and forward networks, which add at least $100$\% latency for each additional hop. 

\begin{figure}[!ht]
\centering
\includegraphics[width=2.5in]{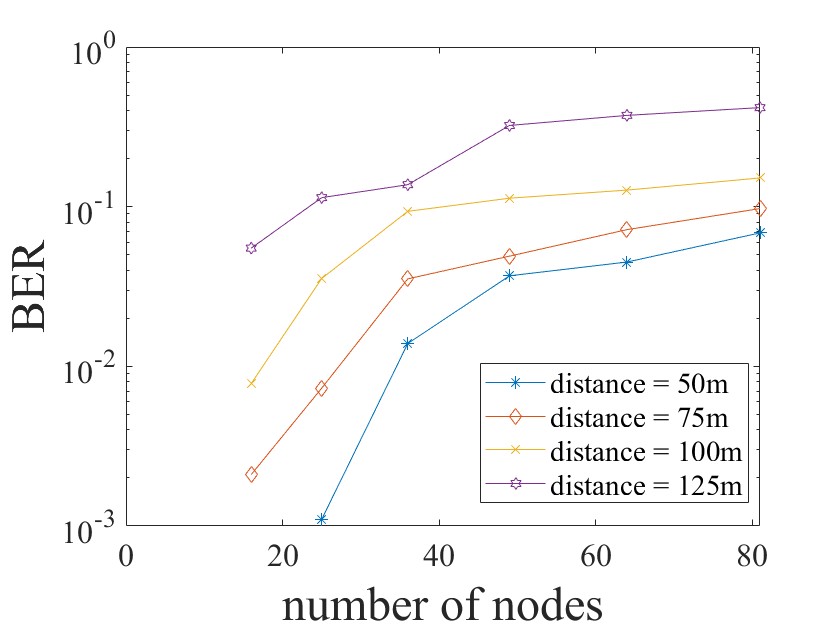}
\caption{Average BER for various numbers of nodes.}
\label{pic9}
\end{figure}

After exploring the distance effect on network BER, we are also concerned about the effect of the number of nodes on network reliability in terms of BER. In this experiment, we fix the grid distance and change the number of nodes instead. Specifically, we first fix the grid distance to $50$~m and transmit $100$ $64$-bit packets through the network consisting of $\{16, 25, 36, 49, 64, 81\}$ nodes. Subsequently, we fix the grid distance to $\{75, 100, 125\}$~m, and repeat the above experiment. The results are illustrated in Figure~\ref{pic9}. According to Figure~\ref{pic9}, we can conclude that the increase in the number of nodes harms network reliability. This is due to the fact that a more dense deployment results in more nodes relaying the same symbol simultaneously, which can cause destructive interference. We leave addressing this problem for future work. Despite these negative effects, our network can still achieve reasonable BER below $5$\% when the number of nodes is less than $50$ and the grid distance is less than $75$~m. 

\subsection{Performance comparasion}
In this section, we compare our network performance with other state-of-the-art low-latency networks. In~\cite{oostvogels2020zero}, performance evaluation shows that Zero-Wire can achieve $99$\% reliability for $32$-bit frame reception. In~\cite{sutton2015zippy}, Zippy reaches $94.6\%$ for $16$-bit frame reception. According to our results, the proposed design with a grid distance of $60$~m can achieve BER of $0.04$\%, which results in $(1-0.0004)^{32} \approx 98.7$\% frame delivery ratio for a $32$-bit frame based on an assumption of statistical independence of subsequent demodulation attempts. Although the reliability performance is similar, the data rate for the proposed design is significantly higher, $100$~kbps in comparison to $20$~kbps supported by Zero-Wire and $1.364$~kbps for Zippy. Besides, a $2.4$~GHz frequency band provides better coverage and propagation properties, allowing reliable data transmission over larger distances.

However, it should be noted that in contrast to~\cite{oostvogels2020zero} and~\cite{sutton2015zippy}, our solution is only tested using simulation and lacks convincable performance evaluation based on real hardware. However, our experiment results demonstrate for the first time how the order of magnitude improvements are possible by properly designed hardware since traditionally empirical platforms face considerable incidental limitations in their implementation. In our future work, we will conduct such experiments using a software-defined radio~(SDR) for a fairer comparison.

\section{Conclusion}
\balance
\label{sec:conclusion}
In this paper, we have proposed an original symbol-synchronous radio design that aims at providing reliable low-latency communication in challenging WSN scenarios, such as industrial IoT, or robotic swarms. In contrast to traditional WSN approaches, the proposed scheme utilizes concurrent transmissions and enables symbol-by-symbol relaying, thus providing wire-like delays through a wireless medium. Our early results show that the proposed design has extremely high potential. With a low transmission power of $0$~dBm, it allows transmitting the signals with BER of $0.04$\% through in a 4x4 node network topology over a range of $250$~m from the initiator to the furthest receiver~(i.e., a grid distance of $60$~m). In addition, in a 7-by-7 node network topology with a grid distance of $75$~m, the current design still achieves to transmit data with BER of $5$\%. As for latency, it can achieve an end-to-end latency of about $5$ milliseconds for a $512$-bit packet and about $0.64$ milliseconds for a $64$-bit packet. Furthermore, while store-and-forward networks scale linearly with the number of hops in terms of latency, our results have shown significantly better scaling behavior, with a latency increase of only $0.3$\% per hop. 

It should be noted that there is a trade-off between the data rate and the experienced bit error rate. A higher data rate implies lower symbol duration, which may result in inter-symbol interference for networks with a large number of nodes. The optimal selection of symbol duration will be considered in our future work.  
Besides that, we plan to improve the reliability of the system by incorporating adaptive transmit power, retransmissions, and error-correcting codes.
Finally, more realistic scenarios for performance evaluation will be considered, such as scenarios with mobile nodes and more complex environments, e.g., modern industrial environments. We aim to validate the proposed approach in a hardware testbed using an SDR-based prototype. 

\section*{Acknowledgment}
This work was supported in part by the European Union’s Horizon Europe Framework under Grant 101093046 - project OpenSwarm, and in part by the Flemish Government under FWO Project G019722N - LOCUSTS. Jonathan Oostvogels is funded by the Research Foundation - Flanders~(FWO), grant number 11H7923N.

\bibliographystyle{IEEEtran}
\bibliography{reference}

\begin{thebibliography}{10}
\providecommand{\url}[1]{#1}
\csname url@samestyle\endcsname
\providecommand{\newblock}{\relax}
\providecommand{\bibinfo}[2]{#2}
\providecommand{\BIBentrySTDinterwordspacing}{\spaceskip=0pt\relax}
\providecommand{\BIBentryALTinterwordstretchfactor}{4}
\providecommand{\BIBentryALTinterwordspacing}{\spaceskip=\fontdimen2\font plus
\BIBentryALTinterwordstretchfactor\fontdimen3\font minus
  \fontdimen4\font\relax}
\providecommand{\BIBforeignlanguage}[2]{{%
\expandafter\ifx\csname l@#1\endcsname\relax
\typeout{** WARNING: IEEEtran.bst: No hyphenation pattern has been}%
\typeout{** loaded for the language `#1'. Using the pattern for}%
\typeout{** the default language instead.}%
\else
\language=\csname l@#1\endcsname
\fi
#2}}
\providecommand{\BIBdecl}{\relax}
\BIBdecl

\bibitem{10.1007/978-3-030-29407-6_20}
A.~Sharma and R.~Sharma, ``{A Review of Applications, Approaches, and
  Challenges in Internet of Things (IoT)},'' in \emph{Proceedings of ICRIC
  2019}, P.~K. Singh, A.~K. Kar, Y.~Singh, M.~H. Kolekar, and S.~Tanwar,
  Eds.\hskip 1em plus 0.5em minus 0.4em\relax Cham: Springer International
  Publishing, 2020, pp. 257--269.

\bibitem{electronics8090981}
\BIBentryALTinterwordspacing
M.~A. Siddiqi, H.~Yu, and J.~Joung, ``{5G Ultra-Reliable Low-Latency
  Communication Implementation Challenges and Operational Issues with IoT
  Devices},'' \emph{Electronics}, vol.~8, no.~9, 2019. [Online]. Available:
  \url{https://www.mdpi.com/2079-9292/8/9/981}
\BIBentrySTDinterwordspacing

\bibitem{gupta2023anchor}
N.~K. Gupta, R.~S. Yadav, R.~K. Nagaria, D.~Gupta, A.~M. Tripathi, and O.~J.
  Pandey, ``{Anchor-Based Void Detouring Routing Protocol in Three Dimensional
  IoT Networks},'' \emph{Computer Networks}, vol. 227, p. 109691, 2023.

\bibitem{badis2015modeling}
H.~Badis and A.~Rachedi, ``{Modeling Tools to Evaluate the Performance of
  Wireless Multi-Hop Networks},'' in \emph{Modeling and Simulation of Computer
  Networks and Systems}.\hskip 1em plus 0.5em minus 0.4em\relax Elsevier, 2015,
  pp. 653--682.

\bibitem{coletti2024real}
C.~Coletti and K.~Williams, ``{Real-Time Control Interface for Research and
  Development Using Commercial-Off-The-Shelf (COTS) Mobile Robotics and
  UAVs},'' in \emph{AIAA SCITECH 2024 Forum}, 2024, p. 0235.

\bibitem{7347997}
M.~O. Farooq, C.~J. Sreenan, K.~N. Brown, and T.~Kunz, ``{RPL-based Routing
  Protocols for Multi-Sink Wireless Sensor Networks},'' in \emph{2015 IEEE 11th
  International Conference on Wireless and Mobile Computing, Networking and
  Communications (WiMob)}, 2015, pp. 452--459.

\bibitem{8368987}
A.~Karaagac, J.~Haxhibeqiri, I.~Moerman, and J.~Hoebeke, ``{Time-Critical
  Communication in 6TiSCH Networks},'' in \emph{2018 IEEE Wireless
  Communications and Networking Conference Workshops (WCNCW)}, 2018, pp.
  161--166.

\bibitem{ben2020controller}
N.~M. Ben~Lakhal, O.~Nasri, L.~Adouane, and J.~B. Hadj~Slama, ``{Controller
  Area Network Reliability: Overview of Design Challenges and Safety Related
  Perspectives of Future Transportation Systems},'' \emph{IET Intelligent
  Transport Systems}, vol.~14, no.~13, pp. 1727--1739, 2020.

\bibitem{ferrari2011efficient}
F.~Ferrari, M.~Zimmerling, L.~Thiele, and O.~Saukh, ``{Efficient Network
  Flooding and Time Synchronization with Glossy},'' in \emph{Proceedings of the
  10th ACM/IEEE International Conference on Information Processing in Sensor
  Networks}.\hskip 1em plus 0.5em minus 0.4em\relax IEEE, 2011, pp. 73--84.

\bibitem{oostvogels2020zero}
J.~Oostvogels, F.~Yang, S.~Michiels, and D.~Hughes, ``{Zero-Wire: a
  Deterministic and Low-Latency Wireless Bus through Symbol-Synchronous
  Transmission of Optical Signals},'' in \emph{Proceedings of the 18th
  Conference on Embedded Networked Sensor Systems}, 2020, pp. 164--178.

\bibitem{zimmerling2020synchronous}
M.~Zimmerling, L.~Mottola, and S.~Santini, ``{Synchronous Transmissions in
  Low-Power Wireless: A Survey of Communication Protocols and Network
  Services},'' \emph{ACM Computing Surveys (CSUR)}, vol.~53, no.~6, pp. 1--39,
  2020.

\bibitem{nahas2021blueflood}
B.~A. Nahas, A.~Escobar-Molero, J.~Klaue, S.~Duquennoy, and O.~Landsiedel,
  ``{BlueFlood: Concurrent Transmissions for Multi-Hop Bluetooth 5—Modeling
  and Evaluation},'' \emph{ACM Transactions on Internet of Things}, vol.~2,
  no.~4, pp. 1--30, 2021.

\bibitem{9312182}
J.~Cheng, P.~Yang, K.~Navaie, Q.~Ni, and H.~Yang, ``A low-latency interference
  coordinated routing for wireless multi-hop networks,'' \emph{IEEE Sensors
  Journal}, vol.~21, no.~6, pp. 8679--8690, 2021.

\bibitem{sutton2015zippy}
F.~Sutton, B.~Buchli, J.~Beutel, and L.~Thiele, ``{Zippy: On-demand Network
  Flooding},'' in \emph{Proceedings of the 13th ACM Conference on Embedded
  Networked Sensor Systems}, 2015, pp. 45--58.

\bibitem{9277601}
A.~W. Mbugua, Y.~Chen, L.~Raschkowski, L.~Thiele, S.~Jaeckel, and W.~Fan,
  ``{Review on Ray Tracing Channel Simulation Accuracy in Sub-6 GHz Outdoor
  Deployment Scenarios},'' \emph{IEEE Open Journal of Antennas and
  Propagation}, vol.~2, pp. 22--37, 2021.

\end{thebibliography}

\end{document}